\documentclass[lettersize,journal]{IEEEtran}
\usepackage{cite}
\usepackage{amsmath,amssymb,amsfonts}
\usepackage{algorithmic}
\usepackage{graphicx}
\usepackage{algorithm,algorithmic}
\usepackage{hyperref}
\usepackage{booktabs}
\usepackage{multirow}
\hypersetup{hidelinks}
\usepackage{textcomp}
\usepackage{xcolor}
\usepackage[caption=false,font=footnotesize]{subfig}
\usepackage[percent]{overpic}

\def\BibTeX{{\rm B\kern-.05em{\sc i\kern-.025em b}\kern-.08em
    T\kern-.1667em\lower.7ex\hbox{E}\kern-.125emX}}
\begin{document}
\title{Edge-centric Brain Transformer: An Edge-centric Functional Connectivity Learning Framework for fMRI-based Brain Disorder Diagnosis}
\author{Dengyi Zhao, Zhiheng Zhou, Mengyao Zhou, Yunping Wang and Xingqin Qi, \IEEEmembership{Member, IEEE}
\thanks{This work was supported in part by the National Natural Science Foundation of China No.12471330, the Shandong Provincial Natural Science Foundation No.ZR2025MS71, the China Postdoctoral Science Foundation No.2026M793388 and the Postdoctoral Innovation Program of Shandong Province No.SDCX-ZG-202603012. (Corresponding author: Xingqin Qi.)}
\thanks{Dengyi Zhao, Zhiheng Zhou,   Yunping Wang and Xingqin Qi are with School of Mathematics and Statistics, Shandong University, Weihai 264209, China (e-mail: zhaodengyi@mail.sdu.edu.cn; zhouzhiheng@amss.ac.cn; 
wangyunping@mail.sdu.edu.cn; 
qixingqin@sdu.edu.cn).}
\thanks{Mengyao Zhou is with Academy of Mathematics and Systems Science, University of Chinese Academy of Sciences, Beijing 100190, China (e-mail: zhoumengyao@amss.ac.cn).}
}

\maketitle
\begin{abstract}
Resting-state functional magnetic resonance imaging (rs-fMRI) enables the characterization of functional interactions among distributed brain regions and has shown promise for brain disorder diagnosis. However, existing deep learning methods predominantly rely on node-centric representations, where brain regions serve as the primary learning units, potentially overlooking discriminative alterations embedded in functional connections. Here, we propose an edge-centric  brain transformer (EBT) framework that reformulates rs-fMRI analysis as functional connection representation learning. Instead of modeling brain regions independently, EBT constructs edge time-series representations to capture dynamic co-fluctuation patterns of functional connections and organizes discriminative connections into a line graph for explicit connection-to-connection modeling. A structure-aware transformer is developed to learn both local dependencies among anatomically related connections and global interactions across distributed functional networks. Furthermore, an edge-level orthogonal clustering readout module is introduced to derive subject-level representations and identify latent connectivity modules associated with brain disorders. Evaluations on multiple neuroimaging datasets demonstrate that EBT consistently outperforms representative graph neural networks, brain transformers, and conventional connectivity-based approaches. Interpretability analyses further reveal stable disease-associated functional connections and connectivity modules that align with known pathological network alterations. These findings establish an edge-centric perspective for rs-fMRI-based brain disorder diagnosis and provide a promising framework for discovering interpretable connectivity biomarkers. The source code is publicly available at:
\url{https://github.com/Zdy12/Edge-centric-Brain-Transformer}.

\end{abstract}


\section{Introduction}
\label{sec:introduction}
\IEEEPARstart{R}{esting-state} functional magnetic resonance imaging (rs-fMRI) provides a non-invasive approach for characterizing spontaneous brain activity and functional interactions among distributed brain regions. By estimating statistical dependencies between regional time series, functional connectivity (FC) networks have become a fundamental framework for investigating large-scale brain organization and identifying abnormal network alterations associated with neurological and psychiatric disorders\cite{bullmore2009complex,smith2013functional,bassett2017network}. Extensive evidence from network neuroscience has demonstrated that brain disorders, including Alzheimer’s disease (AD), autism spectrum disorder (ASD), schizophrenia, and major depressive disorder, are associated with disrupted functional integration, altered modular organization, and abnormal communication patterns across distributed neural systems\cite{menon2011large,dennis2014functional,hull2017resting}. Consequently, rs-fMRI-based computational approaches have attracted increasing attention for disease diagnosis, biomarker discovery, and understanding the mechanisms underlying brain dysfunction.

Traditional FC-based machine learning approaches typically represent each subject using functional connectomes and extract discriminative patterns through handcrafted connectivity features, graph-theoretical measures, or conventional classifiers\cite{richiardi2011decoding,varoquaux2013learning}. These approaches have provided important insights into disease-related functional alterations; however, their performance is often limited by predefined feature representations and the difficulty of capturing nonlinear interactions within high-dimensional connectivity networks. Recent advances in deep learning have provided new opportunities for automatically learning hierarchical representations from brain networks\cite{kan2022brain,bian2023adversarially,tang2025graph,zhao2026hoi}.

Among these approaches, graph neural networks (GNNs) have become a widely adopted framework for modeling brain connectivity networks by representing brain regions as nodes and functional relationships as edges\cite{parisot2018disease,li2021braingnn,tang2025graph}. GNN-based methods naturally incorporate network structure into representation learning and have demonstrated promising performance in brain disorder classification and functional decoding. More recently, transformer-based architectures have also been introduced into neuroimaging analysis due to their capability of modeling long-range dependencies beyond the local aggregation mechanism of conventional GNNs\cite{kan2022brain,yu2024long,wang2026graph}. Despite these advances, most existing deep learning approaches remain fundamentally node-centric, where brain regions are considered the primary computational units and functional connections mainly serve as pathways for information propagation.

Although node-centric learning has been effective for capturing regional characteristics and local neighborhood relationships, it may overlook important pathological information encoded directly in functional connections. Increasing evidence from network neuroscience suggests that brain dysfunction is not solely driven by abnormalities within individual regions but also emerges from disrupted interactions among distributed brain systems\cite{bullmore2009complex,bassett2017network}. Functional connections reflect coordinated neural communication, and alterations in connectivity strength, synchronization patterns, and inter-network interactions have been consistently associated with neurological and psychiatric disorders\cite{greicius2004default,zalesky2010network,menon2011large,faskowitz2020edge,zamani2020high}. Therefore, functional connections themselves may serve as informative biomarkers that capture disease-related alterations beyond regional representations.

Motivated by this observation, edge-centric analysis has recently emerged as a complementary perspective that treats functional connections as fundamental computational units rather than merely relationships between brain regions. Previous neuroscience studies have shown that edge-level representations can reveal transient co-fluctuation patterns and connectivity motifs that are not fully captured by static FC matrices\cite{zamani2020high}. In particular, edge time-series analysis provides a direct characterization of dynamic interactions among functional connections and has demonstrated additional information for studying individual differences and cognitive states\cite{faskowitz2020edge}. Furthermore, graph transformations such as line graph representations provide a natural mathematical framework for converting edge interactions into explicit relational structures, where functional connections can be modeled as nodes and their interactions can be directly learned\cite{evans2009line}.

However, existing edge-centric approaches remain insufficient for rs-fMRI-based brain disorder diagnosis. First, many studies rely on static connectivity representations and neglect the temporal co-fluctuation patterns inherent in fMRI signals. Second, interactions among functional connections are often predefined using handcrafted rules, limiting the ability to adaptively learn complex local and global connectivity dependencies. Third, existing neuroimaging models still lack interpretable mechanisms for identifying disease-associated connectivity modules at the connection level. Therefore, an effective framework that directly learns functional connection representations, captures their interactions, and provides interpretable connectivity biomarkers remains an important challenge.

To address these limitations, we propose an Edge-centric  Brain Transformer (EBT) framework that reformulates rs-fMRI-based brain disorder diagnosis as a functional connection representation learning problem. Instead of treating brain regions as the primary learning units, EBT constructs edge time-series representations to preserve dynamic functional co-fluctuation patterns and organizes discriminative functional connections into a line graph for explicit connection-to-connection modeling. A structure-aware edge transformer is further developed to capture both local dependencies among anatomically related connections and global interactions across distributed functional networks. Moreover, an edge-level orthogonal clustering readout module is introduced to generate subject-level representations while identifying latent connectivity modules associated with brain disorders.

The major contributions of this study are summarized as follows:
\begin{itemize}
    \item We introduce an edge-centric representation learning paradigm for rs-fMRI-based brain disorder diagnosis, where functional connections rather than brain regions serve as the fundamental learning units.
    \item We develop a structure-aware edge transformer based on line graph modeling, which explicitly captures interactions among functional connections by integrating local anatomical dependencies and global network-level relationships.
    \item We propose an interpretable edge-level orthogonal clustering readout module that derives subject-level representations and identifies latent connectivity modules associated with brain disorders.
    \item Comprehensive experiments on multiple neuroimaging datasets demonstrate the effectiveness of EBT, showing superior performance compared with representative GNNs, brain transformers, and conventional FC-based machine learning methods, while revealing biologically meaningful disease-associated connectivity patterns.
\end{itemize}

\subsection{Functional Connectivity-based Brain Disorder Diagnosis}
Functional connectivity (FC) derived from resting-state functional magnetic resonance imaging (rs-fMRI) has been widely used to characterize large-scale functional organization of the human brain and identify abnormal connectivity patterns associated with neurological and psychiatric disorders. Functional connectomes provide a comprehensive representation of distributed neural interactions and have been extensively investigated as potential biomarkers for brain disorders. Bullmore et al. \cite{bullmore2009complex} introduced the complex network perspective for studying structural and functional brain organization, establishing graph-based analysis as a fundamental framework for connectome research. Greicius et al. \cite{greicius2004default} demonstrated that Alzheimer's disease is associated with disrupted default mode network connectivity, highlighting the clinical relevance of resting-state connectivity alterations. Furthermore, network-based statistical analysis revealed that disease-related abnormalities often emerge as distributed connectivity subnetworks rather than isolated connections\cite{zalesky2010network}.

Traditional FC-based machine learning approaches typically represent each subject using connectivity matrices and identify discriminative patterns through handcrafted connectivity features, graph-theoretical measures, or conventional classifiers such as support vector machines and random forests. Richiardi et al. \cite{richiardi2011decoding} investigated functional connectivity-based decoding and demonstrated that connectivity patterns contain predictive information for distinguishing brain states. Varoquaux et al. \cite{varoquaux2013learning} further highlighted the challenges of learning functional connectomes due to high dimensionality, limited sample sizes, and inter-subject variability, issues that remain critical even for modern deep learning approaches. In addition, connectome-based predictive modeling (CPM) has provided a systematic framework for identifying predictive functional connections associated with individual behaviors and clinical phenotypes\cite{shen2017using}. Despite their success, these approaches generally rely on predefined feature extraction or shallow representations, limiting their ability to capture nonlinear interactions among functional connections and hierarchical organization patterns within large-scale connectomes.

\subsection{GNNs for Brain Network Representation Learning}
Graph neural networks (GNNs) have emerged as powerful frameworks for learning representations from brain connectivity networks by explicitly incorporating graph structures into deep learning models. Unlike conventional neural networks designed for Euclidean data, GNNs naturally operate on irregular graph domains, enabling brain regions and their interactions to be modeled as nodes and edges. Early advances in graph deep learning, including graph convolutional networks and graph attention networks, established topology-aware representation learning by aggregating information from neighboring nodes through learnable propagation mechanisms \cite{kipf2016semi,velivckovic2017graph}.

In neuroimaging analysis, GNNs have increasingly been applied to functional and structural brain networks for disease diagnosis and biomarker discovery. Parisot et al.~\cite{parisot2018disease} introduced graph convolutional networks for neurological disorder prediction by constructing population graphs that integrate imaging features with subject-level relationships. Ktena et al. \cite{ktena2017distance} explored metric learning with spectral graph convolutions on brain connectivity networks, demonstrating the effectiveness of graph representation learning for connectome analysis. More recently, interpretable approaches such as BrainGNN have incorporated region-aware pooling mechanisms to identify disease-associated brain regions and improve model interpretability\cite{li2021braingnn}. However, BrainGNN and related approaches primarily operate on node embeddings, where interpretability is achieved by identifying important brain regions rather than directly modeling informative functional connections.

Recent studies have extended GNN-based brain network analysis by incorporating topological information, multi-view learning, and higher-order interactions. Persistent homology-based graph neural networks have been proposed to integrate multiscale topological characteristics into brain connectivity representations, while hypergraph-based models have explored group-wise interactions among distributed brain regions\cite{bian2023adversarially,zhu2025brainchef}. However, these approaches still mainly preserve a node-centric representation paradigm: hypergraph methods model relationships among groups of nodes, whereas functional connections themselves remain implicit. Consequently, direct learning of functional connection representations and interactions among connections remains insufficiently explored, despite the fact that disease-related alterations may emerge from coordinated changes among multiple functional links rather than individual regions.

\subsection{Transformer-based Brain Network Modeling}
Transformer architectures have recently attracted increasing attention in brain network analysis due to their capability of modeling long-range dependencies through self-attention mechanisms. Unlike conventional graph neural networks that rely on local neighborhood aggregation, transformers can directly capture global relationships among input tokens and dynamically assign importance to different network components. Early studies introduced transformer-based frameworks for functional connectivity analysis and demonstrated their potential for learning complex dependencies from brain networks. For example, Brain Network Transformer (BNT) incorporated transformer architectures into functional connectivity analysis and introduced a clustering-based readout mechanism for brain disorder identification\cite{kan2022brain}. Other studies further explored graph transformer architectures and attention-based learning strategies to integrate anatomical constraints and functional connectivity information for neuroimaging prediction tasks\cite{ying2021transformers,qu2023interpretable}.

Recent advances have expanded transformer-based brain network modeling toward more flexible and scalable representation learning. Gated graph transformer frameworks have been proposed to combine graph structures with self-attention mechanisms for functional brain network analysis and cognitive prediction\cite{qu2023interpretable}. Contrastive transformer models have further improved robustness against population heterogeneity and enhanced disease-related representation learning from functional connectomes\cite{xu2024contrasformer}. More recently, transformer-based models have explored sparse attention mechanisms, efficient architectures, and large-scale functional connectome representation learning\cite{li2025miniformer,zhou2025gsaformer,wang2026graph}.

Despite architectural differences, existing GNN and transformer-based brain network models share a common node-centric representational paradigm, where brain regions serve as fundamental learning units and functional connections mainly provide structural constraints or information propagation pathways. Although BNT introduced clustering-based readout for learning functional organization, its representation remains centered on brain regions rather than functional connections. Therefore, directly treating functional connections as learning tokens and modeling interactions among connections represents an important yet underexplored direction for brain network representation learning.

\subsection{Edge-centric Functional Network Representation Learning}
Edge-centric functional network analysis has recently emerged as an alternative perspective that shifts the fundamental representation unit of brain networks from regions to functional connections. Unlike conventional node-centric approaches, where connections are mainly regarded as relationships for information propagation between brain regions, edge-centric frameworks explicitly characterize interactions among functional connections. Faskowitz et al. \cite{faskowitz2020edge} introduced edge-centric functional network representations and demonstrated that edge time-series can reveal overlapping functional organizations beyond traditional node-based connectivity analysis. Zamani Esfahlani  et al. \cite{zamani2020high} further showed that high-amplitude co-fluctuation events captured by edge time-series play an important role in shaping functional connectivity patterns and provide fine-grained information about dynamic brain organization. Mathematical analyses of edge-centric functional connectivity have further established the theoretical basis of edge time-series and highlighted their ability to characterize temporal variations in functional interactions\cite{novelli2022mathematical}.

Recent studies have extended edge-centric representations toward disease characterization and biomarker discovery. Edge time-series based approaches have revealed disease-related alterations that are not fully captured by conventional static functional connectivity. For example, edge-centric network analyses have identified abnormal functional integration patterns in mild cognitive impairment and Alzheimer's disease by characterizing dynamic interactions among functional connections\cite{wang2023edge,chumin2024edge}. Moreover, edge-centric approaches have been applied to investigate altered functional organization in psychiatric disorders, demonstrating their potential for identifying clinically relevant connectivity patterns\cite{esfahlani2022edge,pu2025edge,qin2025edge}. These studies suggest that functional connections themselves contain discriminative information associated with brain dysfunction and may provide complementary biomarkers beyond regional representations.

Despite these advances, existing edge-centric studies mainly focus on descriptive characterization of functional organization, dynamic analysis, or statistical identification of connectivity patterns. Recent computational studies have begun exploring edge-aware representation learning by integrating edge information into graph learning frameworks, aiming to overcome the limitations of conventional node-based models\cite{yang2025edge,santoro2025nodes,merritt2026dual}. However, a unified deep learning framework that directly treats functional connections as learning units, explicitly models interactions among connections, and provides interpretable disease-associated connectivity modules remains largely unexplored. This motivates the development of an edge-centric brain transformer framework for rs-fMRI-based brain disorder diagnosis.


\begin{figure*}[htbp]
  \centering
\includegraphics[scale=0.50]{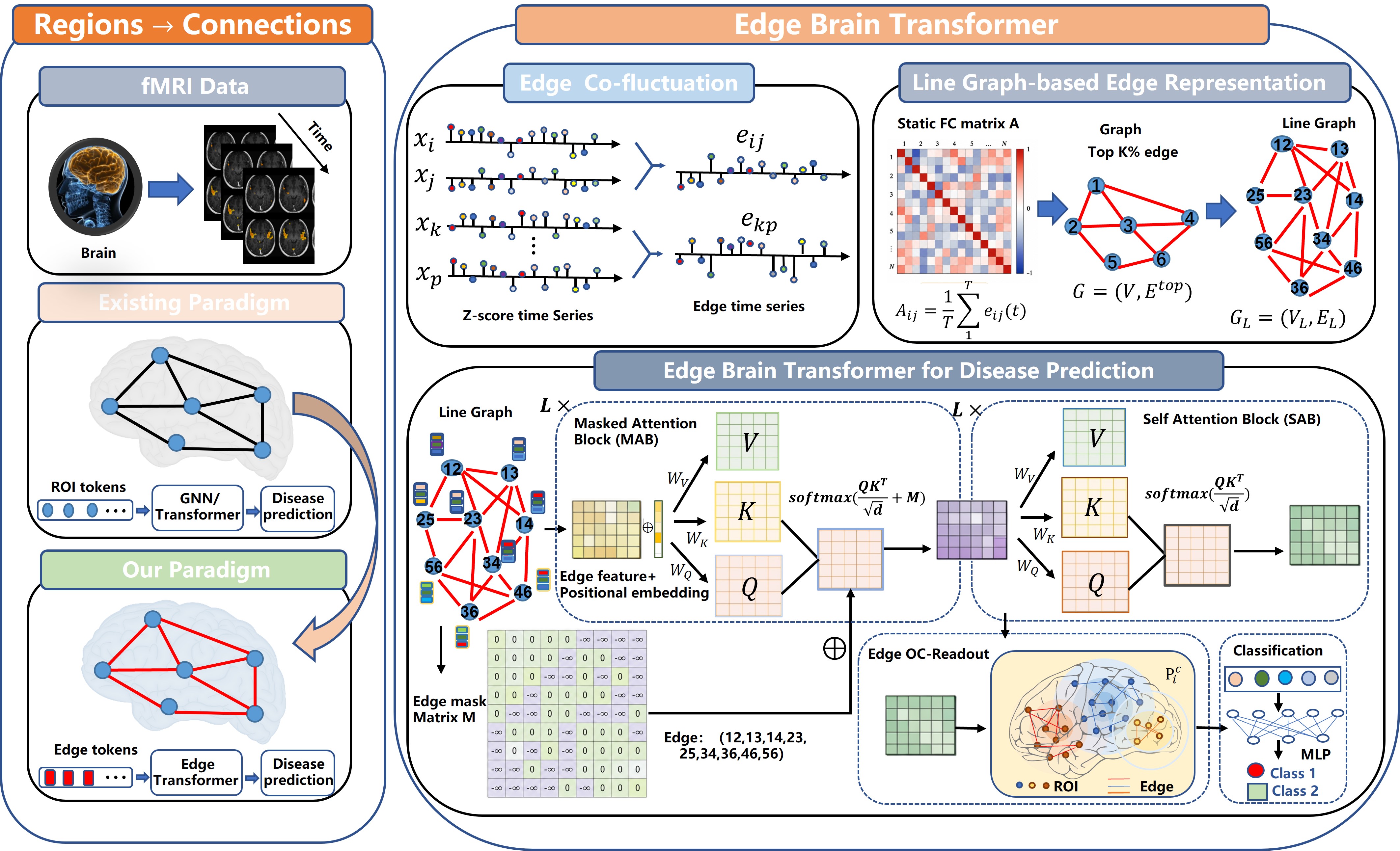}
  \caption{
Overview of the proposed Edge-centric Brain Transformer (EBT) framework for 
fMRI-based brain disorder diagnosis. 
Unlike existing region-centric approaches that represent brain regions as 
fundamental learning units, EBT reformulates brain network learning from 
regions to functional connections. 
First, regional BOLD time series are transformed into edge-wise temporal 
co-fluctuation representations to capture dynamic interactions between 
brain regions beyond static functional connectivity. 
Discriminative functional connections are then selected and converted into 
a line graph, where functional connections are represented as graph nodes 
and their interactions are explicitly modeled. 
A structure-aware edge transformer is subsequently employed to learn both 
local dependencies among anatomically related connections through masked 
attention and global dependencies among distributed functional connections 
through self-attention. 
Finally, an edge-level orthogonal clustering readout (Edge-OCRead) module 
generates subject-level connectivity representations for disease 
classification while identifying interpretable connectivity modules. 
}
  \label{Fig1}
\end{figure*}

\section{Proposed Method}
\label{PROPOSED METHOD}

\subsection{Overview}
Inspired by recent developments in end-to-end attention-based graph learning\cite{buterez2025end}, we propose an edge-centric brain transformer framework for functional MRI-based brain disorder diagnosis. As illustrated in Fig.~\ref{Fig1}, unlike conventional brain network models that represent brain regions as the fundamental computational units, our framework considers functional connections as the primary learning objects. This design is motivated by the observation that neurological disorders are often characterized by distributed alterations of inter-regional connectivity rather than isolated abnormalities of individual brain regions.

The proposed framework consists of four main components:
\begin{itemize}
    \item Constructing edge-wise temporal co-fluctuation representations from fMRI time series via the element-wise product of $z$-scored regional signals.
    
    \item Identifying discriminative functional connections within the training data based on their statistical relevance to diagnostic labels.
    
    \item Learning edge representations through a structure-aware transformer encoder based on line graph modeling, where nodes represent functional connections and graph edges encode shared original brain regions.
    
    \item Introducing an edge-level orthogonal clustering readout that clusters edge embeddings into \(P\) connectivity prototypes with attention-based aggregation for subject-level prediction.
\end{itemize}

Given an fMRI signal matrix \(\mathbf X\), the overall mapping of the proposed framework is formulated as
\begin{equation}
\mathbf X
\rightarrow
\mathcal{E}
\rightarrow
\mathcal{E}^{*}
\rightarrow
G_L
\rightarrow
\mathbf  Z
\rightarrow
\mathbf g_E
\rightarrow
\hat y ,
\end{equation}
where \(\mathbf X\) denotes the original fMRI time-series data, \(\mathcal{E}\) represents the complete set of functional connections, \(\mathcal{E}^{*}\) denotes the selected discriminative connections, \(G_L\) represents the constructed line graph, \(\mathbf Z\) denotes the learned edge representations, \(g_E\) is the subject-level representation, and \(\hat y\) is the predicted disease probability.

\subsection{Edge-wise Temporal Co-fluctuation Representation}
For each subject, the preprocessed resting-state fMRI data are represented as \(\mathbf X=[\mathbf x_1,\mathbf x_2,\ldots,\mathbf x_N]^T\in\mathbb{R}^{N\times T},\) where \(N\) denotes the number of brain regions and \(T\) represents the number of temporal samples. The vector \(\mathbf x_i\in\mathbb{R}^{T}\) denotes the BOLD signal of the \(i\)-th region.

Traditional functional connectivity summarizes the temporal relationship between two regions using a single correlation coefficient. However, this temporal aggregation may obscure transient connectivity alterations. Therefore, we construct edge time series to explicitly characterize temporal co-fluctuation patterns.

For each pair of brain regions, the edge time series is defined as
\begin{equation}
e_{ij}(t)
=
\tilde{x}_{i}(t)
\tilde{x}_{j}(t),
\quad
t=1,\ldots,T ,
\end{equation}
where \(\tilde{x}_{i}(t)\) and \(\tilde{x}_{j}(t)\) denote standardized BOLD signals of regions \(i\) and \(j\), respectively. The edge representation is obtained as
\begin{equation}
\mathbf E_{ij}
=
[e_{ij}(1),e_{ij}(2),...,e_{ij}(T)]
\in
\mathbb{R}^{T},
\end{equation}
where \(\mathbf E_{ij}\) describes the dynamic co-fluctuation pattern of the functional connection between regions \(i\) and \(j\). The conventional static functional connectivity can be recovered as
\begin{equation}
FC_{ij}
=
\frac{1}{T}
\sum_{t=1}^{T}e_{ij}(t),
\end{equation}
where \(FC_{ij}\) denotes the Pearson correlation coefficient between two brain regions. Thus, edge time series can be regarded as a temporal expansion of static functional connectivity while preserving transient connectivity variations.

\subsection{Training-fold-specific Discriminative Edge Selection}
For a brain atlas containing \(N\) regions, the number of possible undirected functional connections is \(M=\frac{N(N-1)}{2},\) where \(M\) represents the total number of candidate functional edges. Directly modeling all edges introduces substantial computational and memory burdens due to the quadratic growth of edge interactions. Therefore, we identify disease-relevant functional connections before edge transformer encoding.

To avoid information leakage, edge selection is independently performed within each training fold. For each edge \((i,j)\), the connectivity strength is defined as \(w_{ij}=FC_{ij},\) where \(w_{ij}\) denotes the functional connectivity value of edge \((i,j)\). For datasets containing multiple diagnostic categories (e.g., CN, MCI, and AD in ADNI), statistical differences among groups are evaluated using a multi-group hypothesis test:
\begin{equation}
p_{ij}
=
\mathcal{T}
(
w_{ij}^{1},
w_{ij}^{2},
\ldots,
w_{ij}^{G}
),
\end{equation}
where \(G\) denotes the number of diagnostic groups, \(w_{ij}^{g}\) represents the connectivity values of edge \((i,j)\) in the \(g\)-th group, and \(\mathcal{T}(\cdot)\) denotes the statistical test, implemented using one-way ANOVA or the Kruskal--Wallis test depending on the data distribution. The candidate edges are ranked according to their statistical significance:
\begin{equation}
\Omega
=
\operatorname{Rank}
(\{p_{ij}\}_{i<j}),
\end{equation}
where \(\Omega\) denotes the ordered edge list. The top \(\rho\) proportion of discriminative edges is retained:
\begin{equation}
\mathcal{E}^{*}
=
\operatorname{Top}^{\lfloor \rho |\Omega| \rfloor}
(\Omega),
\end{equation}
where \(\rho\in(0,1)\) denotes the proportion of selected functional connections and \(|\Omega|\) denotes the total number of candidate edges.

\subsection{Line Graph-based Edge Representation}
After edge selection, the functional connectivity network is converted into a line graph to explicitly characterize interactions among functional connections. Line graph transformation enables explicit modeling of edge-to-edge interactions that are only implicitly encoded in conventional node-centric GNNs.

Given the selected functional network \(G=(V,\mathcal{E}^{*}),\) where $V$ denotes the set of $N$ brain regions. The line graph corresponding to G is defined as \(G_L=(V_L,E_L),\) where each node in \(V_L\) corresponds to one selected functional connection. Specifically, \(V_L=\{v_{ij}|(i,j)\in\mathcal{E}^{*}\},\) where \(v_{ij}\) represents the line-graph node corresponding to functional connection \((i,j)\). The adjacency matrix of the line graph is defined as
\begin{equation}
\mathbf A_L^{(ij)(kl)}
=
\begin{cases}
1,
&
\{i,j\}\cap\{k,l\}\neq\emptyset,\\
0,
&
otherwise,
\end{cases}
\end{equation}
where \(\mathbf A_L\) describes edge-to-edge interactions and two edge nodes are connected when their corresponding functional connections share one brain region.

\subsection{Structure-aware Edge Transformer}
Each selected functional connection is regarded as an edge token. The initial edge representation is defined as
\begin{equation}
\mathbf h_{ij}^{0}
=
\phi
(
\mathbf E_{ij}
\Vert
p_i
\Vert
p_j
),
\end{equation}
where \(\mathbf E_{ij}\) denotes the edge time-series representation, \(p_i\) and \(p_j\) represent learnable positional embeddings of the two endpoint brain regions, \(\Vert\) denotes feature concatenation, and \(\phi(\cdot)\) is a learnable projection function. The initial edge feature matrix is represented as
\begin{equation}
\mathbf H^{0}
=
[\mathbf h_1^{0},\mathbf h_2^{0},\ldots,\mathbf h_{M^*}^{0}]^{T}
\in
\mathbb{R}^{{M^*}\times d},
\end{equation}
where \(M^*\) denotes the number of selected functional connections and \(d\) represents the embedding dimension. At the \(l\)-th transformer layer, the edge representations are projected into query, key, and value matrices:
\begin{equation}
\mathbf Q^{l}=\mathbf H^{l}\mathbf W_Q^{l},
\quad
\mathbf K^{l}=\mathbf H^{l}\mathbf W_K^{l},
\quad
\mathbf V^{l}=\mathbf H^{l}\mathbf W_V^{l},
\end{equation}
where \(\mathbf W_Q^{l}\), \(\mathbf W_K^{l}\), and \(\mathbf W_V^{l}\) are learnable projection matrices, and \(\mathbf Q^{l}\), \(\mathbf K^{l}\), and \(\mathbf V^{l}\) represent query, key, and value matrices, respectively.

To preserve the intrinsic topology among functional connections, we employ masked edge attention\cite{buterez2025end}. The masked attention restricts the initial local interaction to structurally related edges, providing a topology-aware inductive bias before global integration. The attention mask is constructed according to the line graph adjacency:
\begin{equation}
\mathbf M_{ab}
=
\begin{cases}
0,
&
\mathbf A_L^{ab}=1,\\
-\infty,
&
\mathbf A_L^{ab}=0,
\end{cases}
\end{equation}
where \(\mathbf A_L\) denotes the line graph adjacency matrix and \(M\) restricts information propagation only between structurally related edge tokens.

The local edge interaction is calculated as
\begin{equation}
\widetilde{\mathbf{H}}^{l+1}
=
\operatorname{MAB}
(\mathbf H^{l})
=
\operatorname{softmax}
\left(
\frac{\mathbf Q^{l}(\mathbf K^{l})^{T}}
{\sqrt d}
+
\mathbf M
\right)
\mathbf V^{l},
\end{equation}
where \(\widetilde{\mathbf H}^{l+1}\) denotes the intermediate edge representation after topology-aware attention. Subsequently, a global self-attention operation is applied to capture long-range dependencies among functional connections:
\begin{equation}
\mathbf H^{l+1}
=
\operatorname{SAB}
(\widetilde{\mathbf H}^{l+1})
=
\operatorname{softmax}
\left(
\frac{\widetilde{\mathbf Q}^{l}
(\widetilde{\mathbf K}^{l})^{T}}
{\sqrt d}
\right)
\widetilde{\mathbf V}^{l},
\end{equation}
where \(\widetilde{\mathbf{Q}}^{l}\), \(\widetilde{\mathbf K}^{l}\), and \(\widetilde{\mathbf V}^{l}\) are generated from \(\widetilde{\mathbf H}^{l+1}\) through independent linear projections.

After stacking (L) edge transformer layers, an outer residual connection is introduced to preserve the edge-specific information encoded in the initial representations. The final edge representations are obtained as
\begin{equation}
\mathbf{Z}=\mathbf{H}^{L}+\mathbf{H}^{0},
\end{equation}
where \(\mathbf{H}^{0}\) and \(\mathbf{H}^{L}\) denote the initial edge representations and the output of the (L)-th edge transformer layer, respectively, and \(\mathbf Z\in\mathbb{R}^{M^*\times d}\) denotes the resulting edge embeddings.
\subsection{Edge-level Orthonormal Clustering Readout}

To generate subject-level representations, we proposed an edge-level orthogonal clustering readout module inspired by Brain Network Transformer
~\cite{kan2022brain} from brain-region nodes to functional connection
tokens. The module exploits the assumption that functional connections
with similar learned representations can be softly grouped into latent
connectivity modules.

Let
\(\mathbf Z=[\mathbf z_1,\mathbf z_2,\ldots,\mathbf z_{M^*}]^T
\in\mathbb{R}^{M^*\times d}\)
denote the learned representations of the selected functional
connections. Given \(P\) learnable cluster centers, represented by
\[
\mathbf W_c
=
[\mathbf w_1,\mathbf w_2,\ldots,\mathbf w_P]
\in\mathbb{R}^{d\times P},
\]
the probability of assigning the \(k\)-th functional connection to the
\(p\)-th latent connectivity module is calculated using a softmax
projection:
\begin{equation}
C_{kp}
=
\frac{
\exp\left(\langle \mathbf z_k,\mathbf w_p\rangle\right)
}{
\sum_{q=1}^{P}
\exp\left(\langle \mathbf z_k,\mathbf w_q\rangle\right)
},
\end{equation}
where \(\langle\cdot,\cdot\rangle\) denotes the inner product. The
resulting soft assignment matrix can be written as
\begin{equation}
\mathbf C
=
\operatorname{softmax}
\left(
\mathbf Z\mathbf W_c
\right)
\in\mathbb{R}^{M^*\times P},
\end{equation}
where the softmax operation is applied across the \(P\) latent
connectivity modules for each functional connection.

The selected functional connections are aggregated according to their
soft assignments:
\begin{equation}
\mathbf S
=
\mathbf C^T\mathbf Z
\in\mathbb{R}^{P\times d},
\end{equation}
where each row of \(\mathbf S\) represents the aggregated representation
of one latent connectivity module.
Following the original Orthonormal Clustering Readout, the cluster
centers are initialized as orthonormal bases. 

Although orthonormal initialization provides well-separated initial
cluster centers, the centers and edge assignments are updated during
end-to-end optimization. To reduce redundancy among the learned
connectivity modules and prevent degenerate edge assignments, we further
introduce a normalized orthogonality regularization term on the soft
assignment matrix:
\begin{equation}
\mathcal{L}_{\mathrm{orth}}
=
\left\|
\frac{
\mathbf C^{T}\mathbf C
}{
\left\|
\mathbf C^{T}\mathbf C
\right\|_{F}
+
\epsilon
}
-
\frac{
\mathbf I_{P}
}{
\sqrt{P}
}
\right\|_{F},
\end{equation}
where \(\mathbf I_{P}\) denotes the \(P\)-dimensional identity matrix
and \(\epsilon\) is a small constant introduced for numerical stability.
The normalization removes the dependence of the regularization scale on
the number of selected functional connections. The off-diagonal terms
penalize overlapping assignments between different connectivity modules,
whereas the diagonal terms discourage degenerate solutions in which most
connections are assigned to only a small subset of modules.

Finally, the latent connectivity-module representations are flattened to
obtain the subject-level edge-centric representation:
\begin{equation}
\mathbf g_E
=
\operatorname{Flatten}
\left(
\mathbf S
\right)
\in\mathbb{R}^{Pd},
\end{equation}
where \(\mathbf g_E\) is subsequently used for disease classification.

\subsection{Classification and Optimization}

The class probability vector of the \(s\)-th subject is predicted using
a multilayer perceptron:
\begin{equation}
\hat{\mathbf y}_s
=
\operatorname{softmax}
\left(
\mathbf W_2
\delta
\left(
\mathbf W_1\mathbf g_E^{(s)}
+
\mathbf b_1
\right)
+
\mathbf b_2
\right),
\end{equation}
where \(\mathbf W_1\), \(\mathbf W_2\), \(\mathbf b_1\), and
\(\mathbf b_2\) are learnable parameters, \(\delta(\cdot)\) denotes the
nonlinear activation function, and
\(\hat{\mathbf y}_s
=
[\hat y_{s1},\hat y_{s2},\ldots,\hat y_{sK}]^T
\in\mathbb{R}^{K}\)
denotes the predicted probability vector over \(K\) diagnostic
categories.

The classification objective is defined using the categorical
cross-entropy loss:
\begin{equation}
\mathcal{L}_{\mathrm{cls}}
=
-\frac{1}{S}
\sum_{s=1}^{S}
\sum_{k=1}^{K}
y_{sk}
\log
\hat y_{sk},
\end{equation}
where \(S\) denotes the number of subjects, \(K\) denotes the number of
diagnostic categories, \(y_{sk}\in\{0,1\}\) indicates whether the
\(s\)-th subject belongs to the \(k\)-th category, and
\(\hat y_{sk}\) denotes the predicted probability that the \(s\)-th
subject belongs to the \(k\)-th category. For the ADNI dataset,
\(K=3\), whereas \(K=2\) for the ADHD, ABIDE I, and ABIDE II datasets.

The final optimization objective combines the classification loss and
the orthogonal clustering regularization:
\begin{equation}
\mathcal{L}
=
\mathcal{L}_{\mathrm{cls}}
+
\lambda
\mathcal{L}_{\mathrm{orth}},
\end{equation}
where \(\lambda\) controls the contribution of the orthogonal clustering
regularization.

\section{Experiments}
\label{EXPERIMENTS}
\begin{table}[ht]
  \centering
  \caption{Class distributions of the evaluated rs-fMRI datasets.}
  \label{tab:class_distribution}
  \scalebox{0.85}{
  \begin{tabular}{@{}lllc@{}}
    \toprule
    Dataset & Class & \# Subjects & Disease type \\
    \midrule

    \multirow{3}{*}{ADNI}
    & AD  & 118
    & \multirow{3}{*}{\parbox{5.0cm}{\centering Alzheimer's disease}} \\
    & MCI & 117 & \\
    & CN  & 76  & \\
    \midrule

    \multirow{2}{*}{ADHD}
    & ADHD & 304
    & \multirow{2}{*}{\parbox{5.0cm}{\centering
    Attention-deficit/hyperactivity disorder}} \\
    & Control & 218 & \\
    \midrule

    \multirow{2}{*}{ABIDE I}
    & ASD & 537
    & \multirow{2}{*}{\parbox{5.0cm}{\centering
    Autism spectrum disorder}} \\
    & Control & 488 & \\
    \midrule

    \multirow{2}{*}{ABIDE II}
    & ASD & 238
    & \multirow{2}{*}{\parbox{5.0cm}{\centering
    Autism spectrum disorder}} \\
    & Control & 238 & \\
    \bottomrule
  \end{tabular}
  }
\end{table}
\subsection{Datasets and Preprocessing}

We evaluated the proposed Edge-centric  Brain Transformer (EBT) on four
resting-state functional magnetic resonance imaging (rs-fMRI) cohorts, covering three representative brain disorders: Alzheimer's disease,
attention-deficit/hyperactivity disorder, and autism spectrum disorder.
Specifically, the experiments were conducted on the ADNI, ADHD, ABIDE I, and ABIDE II datasets. \textbf{1) ADNI~\cite{mueller2005ways}.}
The Alzheimer's Disease Neuroimaging Initiative\footnote{\url{https://adni.loni.usc.edu/}} (ADNI) cohort used in this study consists of 311 subjects, including 118 subjects with Alzheimer's disease (AD), 117 subjects with mild cognitive impairment (MCI), and 76 cognitively normal controls (CN). \textbf{2) ADHD~\cite{adhd2012adhd}.}
The attention-deficit/hyperactivity disorder\footnote{\url{https://fcon-1000.projects.nitrc.org/indi/adhd200/}} (ADHD) cohort contains 522 subjects, including 304 subjects with ADHD and 218 typically developing
controls. \textbf{3) ABIDE I~\cite{di2014autism}.}
The Autism Brain Imaging Data Exchange I\footnote{\url{https://fcon\_1000.projects.nitrc.org/indi/abide/abide\_I.html}} (ABIDE I) cohort contains 1,025 subjects collected from multiple imaging sites, including 537 subjects with autism spectrum disorder (ASD) and 488 healthy controls. Owing to its relatively large sample size and substantial inter-site heterogeneity, ABIDE I provides a challenging setting for evaluating the
robustness of functional brain network models. \textbf{4) ABIDE II~\cite{di2017enhancing}.} The ABIDE II cohort\footnote{\url{https://fcon\_1000.projects.nitrc.org/indi/abide/abide\_II.html}} contains 476 subjects, including 238 subjects
with ASD and 238 healthy controls. ABIDE II was used in two complementary experimental settings. The class distributions of the four datasets are
summarized in Table~\ref{tab:class_distribution}.

For all datasets, the resting-state fMRI data were preprocessed using fMRIPrep following the standardized preprocessing procedure described in ~\cite{esteban2018fmriprep}, which consists of multiple steps, including motion correction, realignment, field unwarping, normalization, bias field correction, and brain extraction.  The preprocessed brain images were parcellated into
\(N\) regions of interest according to The Automated Anatomical Labeling (AAL) atlas ~\cite{tzourio2002automated}. For each
subject, the mean blood-oxygen-level-dependent signal was extracted from
each region and standardized along the temporal dimension. The resulting
regional time series were subsequently used to construct edge-wise
co-fluctuation time series and subject-specific functional connection
representations.

\subsection{Baselines and Evaluation Metrics}

We compared EBT with representative methods from four categories. \textbf{1) Traditional FC-based methods.} MLP and SVM\cite{pedregosa2011scikit} directly use vectorized functional connectivity features for classification. These methods provide important non-graph baselines for examining whether explicit graph representation learning is necessary. \textbf{2) Graph neural networks.}
GCN~\cite{kipf2016semi} and GAT~\cite{velivckovic2017graph} represent general message-passing graph neural networks.
BrainGNN~\cite{li2021braingnn}, FBNetGen~\cite{kan2022fbnetgen}, and FROG~\cite{zhang2025frog} are included as representative brain-network learning methods that incorporate brain-specific graph construction, pooling, or invariant representation mechanisms. \textbf{3) General graph transformers.}
Graph Transformer~\cite{yun2019graph}, SAN~\cite{kreuzer2021rethinking}, and Graphormer~\cite{ying2021transformers} are employed to evaluate whether
general-purpose node-centric transformer architectures can effectively
model functional brain networks. \textbf{4) Brain transformers.}
BNT~\cite{kan2022brain}, ALTER~\cite{yu2024long}, and CAGT~\cite{pei2025community} represent recent transformer architectures
specifically developed for functional brain network analysis. These methods
constitute the most direct competitors to EBT because they also use
attention mechanisms but predominantly treat brain regions as tokens.

All methods were evaluated using accuracy, precision, recall, F1-score,
and the area under the receiver operating characteristic curve (AUC).
The reported results are expressed as percentages and summarized as the
mean and standard deviation across evaluation folds. 
For multiclass diagnosis on ADNI, the macro-averaged one-vs-rest AUC was reported.
\begin{table*}[ht]
  \centering
  \caption{Performance comparison with four categories of baselines on four datasets (\%). The best results are marked in bold.}
  \scalebox{0.75}{
  \begin{tabular}{@{}cc|ccccc|ccccc@{}}
    \toprule
    \multirow{2}{*}{Type}
    & \multirow{2}{*}{Method}
    & \multicolumn{5}{c|}{ADNI}
    & \multicolumn{5}{c}{ADHD} \\
    \cmidrule(lr){3-7}
    \cmidrule(lr){8-12}
    &
    & Accuracy & Precision & Recall & F1-score & AUC
    & Accuracy & Precision & Recall & F1-score & AUC \\
    \midrule

    \multirow{2}{*}{\parbox{3cm}{\centering Traditional FC}}
    & MLP
    & 64.1$\pm$5.4
    & 65.8$\pm$9.1
    & 65.8$\pm$4.6
    & 63.1$\pm$6.9
    & 81.2$\pm$2.0
    & 61.5$\pm$14.2
    & 61.4$\pm$23.0
    & 63.7$\pm$10.3
    & 56.8$\pm$18.5
    & 79.2$\pm$4.3 \\

    & SVM
    & 60.0$\pm$9.4
    & 63.9$\pm$7.8
    & 60.6$\pm$8.7
    & 60.0$\pm$9.3
    & 78.6$\pm$8.3
    & 62.2$\pm$6.0
    & 63.6$\pm$7.3
    & 61.8$\pm$5.1
    & 60.6$\pm$5.1
    & 66.6$\pm$5.5 \\
    \midrule

    \multirow{5}{*}{\parbox{3cm}{\centering Graph Neural Networks}}
    & GCN
    & 64.1$\pm$9.9
    & 67.3$\pm$10.6
    & 64.2$\pm$8.9
    & 63.8$\pm$9.7
    & 80.5$\pm$6.6
    & 53.5$\pm$10.6
    & 45.9$\pm$22.7
    & 56.5$\pm$6.4
    & 47.0$\pm$16.1
    & 58.8$\pm$13.7 \\

    & GAT
    & 69.7$\pm$8.4
    & 72.0$\pm$7.0
    & 70.9$\pm$7.8
    & 69.7$\pm$8.7
    & 84.2$\pm$4.6
    & 61.5$\pm$7.8
    & 60.7$\pm$8.2
    & 60.3$\pm$7.9
    & 60.1$\pm$7.9
    & 64.5$\pm$9.4 \\

    & BrainGNN
    & 74.9$\pm$8.4
    & 76.3$\pm$8.6
    & 74.9$\pm$9.1
    & 74.7$\pm$8.7
    & 89.7$\pm$4.4
    & 68.0$\pm$6.5
    & 67.2$\pm$7.1
    & 67.0$\pm$7.5
    & 66.8$\pm$7.3
    & 72.7$\pm$5.8 \\

    & FBNetGen
    & 72.8$\pm$3.9
    & 74.7$\pm$4.9
    & 73.8$\pm$3.9
    & 73.0$\pm$4.5
    & 88.7$\pm$3.4
    & 74.5$\pm$4.8
    & 76.9$\pm$3.3
    & 72.4$\pm$6.7
    & 71.8$\pm$8.0
   &\textbf{  84.5$\pm$2.5} \\

    & FROG
    & 68.2$\pm$5.6
    & 71.2$\pm$6.2
    & 69.9$\pm$5.4
    & 68.1$\pm$6.0
    & 85.0$\pm$3.7
    & 70.3$\pm$4.6
    & 69.9$\pm$4.3
    & 69.8$\pm$3.9
    & 69.6$\pm$4.3
    & 75.9$\pm$4.6 \\
    \midrule

    \multirow{3}{*}{\parbox{3cm}{\centering Graph Transformers}}
    & Graph Transformer
    & 67.2$\pm$6.6
    & 67.9$\pm$5.4
    & 67.5$\pm$6.6
    & 67.1$\pm$6.4
    & 84.7$\pm$3.2
    & 70.1$\pm$4.8
    & 69.8$\pm$5.4
    & 69.3$\pm$4.8
    & 69.1$\pm$4.7
    & 76.4$\pm$2.6 \\

    & SAN
    & 71.3$\pm$7.8
    & 75.2$\pm$6.2
    & 72.3$\pm$7.0
    & 71.6$\pm$8.1
    & 86.8$\pm$3.0
    & 67.4$\pm$4.1
    & 67.9$\pm$3.6
    & 66.6$\pm$3.5
    & 66.0$\pm$3.9
    & 74.0$\pm$4.6 \\

    & Graphormer
    & 70.3$\pm$5.3
    & 71.8$\pm$4.7
    & 71.2$\pm$4.7
    & 70.3$\pm$5.5
    & 88.2$\pm$2.8
    & 54.6$\pm$7.6
    & 54.5$\pm$6.4
    & 53.0$\pm$3.6
    & 48.3$\pm$8.4
    & 57.5$\pm$7.4 \\
    \midrule

    \multirow{3}{*}{\parbox{3cm}{\centering Brain Transformers}}
    & BNT
    & 74.9$\pm$4.6
    & 76.8$\pm$4.7
    & 75.4$\pm$4.9
    & 74.7$\pm$4.8
    & 90.1$\pm$2.0
    & 76.2$\pm$5.5
    & 75.8$\pm$5.7
    & 75.6$\pm$5.5
    & 75.6$\pm$5.6
    & 82.4$\pm$4.7 \\

    & ALTER
    & 70.8$\pm$5.3
    & 72.4$\pm$4.6
    & 71.7$\pm$4.8
    & 70.8$\pm$5.5
    & 88.2$\pm$2.7
    & 69.9$\pm$2.8
    & 69.7$\pm$2.6
    & 69.2$\pm$2.3
    & 69.0$\pm$2.4
    & 74.6$\pm$2.1 \\

    & CAGT
    & 68.2$\pm$8.8
    & 70.5$\pm$10.4
    & 69.5$\pm$8.5
    & 68.2$\pm$8.8
    & 86.1$\pm$3.4
    & 70.1$\pm$0.7
    & 70.0$\pm$0.9
    & 68.7$\pm$1.6
    & 68.6$\pm$1.3
    & 77.0$\pm$2.0 \\
    \midrule

    Our Framework
    & EBT
    & \textbf{76.4$\pm$4.9}
    & \textbf{78.0$\pm$5.0}
    & \textbf{76.8$\pm$5.4}
    & \textbf{76.3$\pm$5.2}
    & \textbf{90.5$\pm$2.3}
    & \textbf{78.3$\pm$2.4}
    & \textbf{77.8$\pm$2.4}
    & \textbf{77.8$\pm$2.7}
    & \textbf{77.7$\pm$2.6}
    & \textbf{84.5$\pm$1.9} \\
     \toprule
    \multirow{2}{*}{Type}
    & \multirow{2}{*}{Method}
    & \multicolumn{5}{c}{ABIDE I}
    & \multicolumn{5}{c}{ABIDE II} \\
    \cmidrule(lr){3-7}
    \cmidrule(lr){8-12}
    &
    & Accuracy & Precision & Recall & F1-score & AUC
    & Accuracy & Precision & Recall & F1-score & AUC \\
    \midrule

    \multirow{2}{*}{\parbox{3cm}{\centering Traditional FC}}
    & MLP
    & 57.6$\pm$2.1
    & 57.5$\pm$2.1
    & 57.3$\pm$2.1
    & 57.2$\pm$2.2
    & 61.0$\pm$2.7
    & 58.2$\pm$6.8
    & 58.3$\pm$6.7
    & 58.2$\pm$6.7
    & 57.6$\pm$7.5
    & 63.3$\pm$9.6 \\

    & SVM
    & 59.4$\pm$3.9
    & 59.3$\pm$3.9
    & 59.2$\pm$3.9
    & 59.2$\pm$3.9
    & 62.7$\pm$4.5
    & 61.5$\pm$5.2
    & 62.0$\pm$5.3
    & 61.5$\pm$5.2
    & 61.1$\pm$5.3
    & 66.2$\pm$8.2 \\
    \midrule

    \multirow{5}{*}{\parbox{3cm}{\centering Graph Neural Networks}}
    & GCN
    & 55.8$\pm$2.4
    & 55.9$\pm$3.0
    & 55.7$\pm$3.1
    & 54.4$\pm$4.9
    & 58.9$\pm$3.0
    & 51.5$\pm$2.1
    & 37.7$\pm$17.5
    & 51.3$\pm$1.8
    & 38.7$\pm$7.4
    & 55.5$\pm$6.4 \\

    & GAT
    & 59.9$\pm$2.7
    & 60.5$\pm$2.1
    & 59.4$\pm$3.4
    & 58.0$\pm$5.1
    & 62.5$\pm$3.9
    & 61.3$\pm$2.6
    & 61.4$\pm$5.2
    & 61.3$\pm$2.5
    & 61.0$\pm$9.8
    & 61.8$\pm$2.2 \\

    & BrainGNN
    & 58.4$\pm$4.5
    & 58.4$\pm$4.6
    & 58.4$\pm$4.6
    & 58.4$\pm$4.6
    & 62.9$\pm$5.3
    & 71.0$\pm$4.7
    & 71.6$\pm$4.6
    & 71.0$\pm$4.7
    & 70.8$\pm$4.8
    & 79.0$\pm$4.0 \\

    & FBNetGen
    & 62.1$\pm$1.4
    & 62.0$\pm$1.3
    & 62.0$\pm$1.3
    & 62.0$\pm$1.3
    & 67.1$\pm$1.4
    & 65.5$\pm$10.2
    & 67.3$\pm$10.0
    & 65.6$\pm$10.1
    & 64.4$\pm$11.3
    & 71.9$\pm$13.6 \\

    & FROG
    & 62.5$\pm$3.1
    & 62.6$\pm$3.2
    & 62.4$\pm$3.3
    & 62.2$\pm$3.3
    & 67.0$\pm$3.1
    & 59.0$\pm$7.4
    & 59.3$\pm$7.6
    & 59.0$\pm$7.4
    & 58.8$\pm$7.4
    & 61.9$\pm$10.4 \\
    \midrule

    \multirow{3}{*}{\parbox{3cm}{\centering Graph Transformers}}
    & Graph Transformer
    & 61.5$\pm$2.8
    & 61.5$\pm$2.7
    & 61.5$\pm$2.7
    & 61.4$\pm$2.7
    & 67.3$\pm$1.2
    & 53.8$\pm$9.2
    & 53.8$\pm$9.5
    & 53.8$\pm$9.3
    & 53.6$\pm$9.4
    & 57.4$\pm$12.7 \\

    & SAN
    & 62.5$\pm$1.2
    & 62.5$\pm$1.2
    & 62.4$\pm$1.3
    & 62.4$\pm$1.3
    & 67.2$\pm$2.6
    & 58.2$\pm$6.3
    & 58.2$\pm$6.3
    & 58.2$\pm$6.3
    & 58.1$\pm$6.4
    & 62.2$\pm$5.9 \\

    & Graphormer
    & 57.6$\pm$1.1
    & 58.0$\pm$1.0
    & 57.7$\pm$1.2
    & 57.1$\pm$1.6
    & 63.3$\pm$2.1
    & 55.2$\pm$7.4
    & 55.1$\pm$8.2
    & 55.3$\pm$7.4
    & 53.9$\pm$9.1
    & 57.6$\pm$13.1 \\
    \midrule

    \multirow{3}{*}{\parbox{3cm}{\centering Brain Transformers}}
    & BNT
    & 63.1$\pm$2.7
    & 63.3$\pm$3.0
    & 62.9$\pm$2.7
    & 62.8$\pm$2.6
    & 68.0$\pm$3.9
    & 72.3$\pm$4.2
    & 72.6$\pm$4.3
    & 72.3$\pm$4.1
    & 72.2$\pm$4.2
    & \textbf{79.3$\pm$2.8} \\

    & ALTER
    & 62.3$\pm$2.8
    & 62.3$\pm$2.8
    & 62.3$\pm$2.7
    & 62.2$\pm$2.7
    & 67.2$\pm$1.2
    & 67.5$\pm$3.6
    & 67.9$\pm$3.4
    & 67.5$\pm$3.6
    & 67.4$\pm$3.7
    & 74.2$\pm$3.0 \\

    & CAGT
    & 64.3$\pm$2.4
    & 64.4$\pm$2.2
    & 64.4$\pm$2.2
    & 64.2$\pm$2.4
    & \textbf{69.7$\pm$1.9}
    & 60.1$\pm$4.3
    & 60.7$\pm$4.5
    & 60.0$\pm$4.2
    & 59.4$\pm$4.3
    & 62.4$\pm$6.3 \\
    \midrule

    \multirow{1}{*}{Our Framework}
    & EBT
    & \textbf{65.5$\pm$2.4}
    & \textbf{65.5$\pm$2.4}
    & \textbf{65.6$\pm$2.4}
    & \textbf{65.5$\pm$2.4}
    & \textbf{69.7$\pm$2.1}
    & \textbf{73.1$\pm$4.5}
    & \textbf{73.4$\pm$4.5}
    & \textbf{73.1$\pm$4.4}
    & \textbf{73.0$\pm$4.5}
    & \textbf{79.3$\pm$2.7} \\
    \bottomrule
  \end{tabular}
  \label{tab:performance_comparison}
}
\end{table*}
\subsection{Implementation Details }
All experiments were implemented in Python 3.9 and PyTorch 2.4.0 on a single NVIDIA GeForce RTX 4060 Ti GPU. For each subject, the ROI-wise BOLD signals were temporally standardized, and each edge time series was constructed as the pointwise product of the standardized signals of its two endpoint ROIs. Edge selection was performed exclusively within each training fold. All candidate connections were ranked according to the \(p\)-values obtained from one-way ANOVA or the Kruskal--Wallis test, and the top \(\rho\in\{0.2,0.3\}\) proportion of the most statistically discriminative connections was retained. The selected edge indices were then fixed for the corresponding validation and test subjects. Each edge token combined its co-fluctuation representation, ROI embedding and edge-identity embedding. The proposed Edge-centric  Brain Transformer used an embedding dimension of \(d=32\), two encoder blocks arranged as masked edge attention followed by global self-attention,  dropout of 0.2,  and an encoder-level residual connection. The structural mask allowed attention only between edges sharing at least one ROI. Edge-OCRead employed 16 orthogonally regularized latent edge modules for graph-level representation learning. The model was trained using AdamW with a learning rate of \(2\times10^{-4}\), weight decay of \(10^{-3}\) and batch size 32 for at most 20 epochs. Within-cohort performance was evaluated using stratified five-fold cross-validation.

\subsection{Performance Comparison}

The results in Table~\ref{tab:performance_comparison} compare EBT with
traditional FC-based classifiers, GNNs, general graph
transformers, and brain-specific transformer models. EBT consistently
achieves the best performance across all four within-cohort evaluations,
highlighting the advantage of directly learning functional connection
representations over region-level features.

On the ADNI dataset, EBT achieves an accuracy of
\(76.4\pm4.9\%\) and an AUC of \(90.5\pm2.3\%\), outperforming all
competing methods. Compared with BNT, the strongest brain-transformer
baseline on this dataset, EBT improves accuracy, precision, recall,
F1-score, and AUC by 1.5\%, 1.2\%, 1.4\%, 1.6\%, and 0.4\%,
respectively. These improvements indicate that connection-level
representations capture additional discriminative patterns associated with
AD, MCI, and CN classification.

For the ADHD dataset, EBT achieves an accuracy of
\(78.3\pm2.4\%\) and an F1-score of \(77.7\pm2.6\%\), exceeding the best
competing results from BNT by 2.1\% in both metrics.
Moreover, EBT obtains the highest recall of \(77.8\pm2.7\%\), suggesting
that edge-centric modeling can effectively identify disease-related
functional alterations while maintaining stable performance across
evaluation folds.

On the ABIDE I cohort, EBT achieves an accuracy of
\(65.5\pm2.4\%\), improving over CAGT by 1.2\%.
It also provides consistent gains in precision, recall, and F1-score,
while maintaining a comparable AUC of \(69.7\%\). For ABIDE II, EBT
achieves an accuracy of \(73.1\pm4.5\%\) and improves all threshold-based
metrics over BNT by 0.8\%, further demonstrating its
robustness under multi-site autism datasets.

Several observations can be drawn from these comparisons. First,
traditional GNNs and general graph transformers do not consistently
outperform FC-based classifiers, suggesting that conventional node-centric
message passing may lose important connection-level information. Second,
brain-specific transformer models generally achieve better performance,
highlighting the importance of architectures designed for functional
connectome analysis. Finally, EBT consistently surpasses existing
node-centric brain transformers, supporting the hypothesis that
disease-related abnormalities are encoded not only in individual brain
regions but also in interactions among functional connections. Unlike conventional GNNs that aggregate edge information into node
features,
EBT preserves selected functional connections as explicit tokens and
models their interactions through structure-aware edge attention. The
consistent improvements across AD, ADHD, and ASD classification tasks
suggest that edge-wise temporal co-fluctuation representations provide a
more informative and generalizable characterization of distributed brain
network alterations.

\begin{table}[ht]
  \centering
  \caption{Performance comparison with four categories of baselines in external validation from ABIDE I to the  ABIDE II cohort (\%). The best results are marked in bold.}
  \scalebox{0.7}{
  \begin{tabular}{@{}ccccccc@{}}
    \toprule
    \multirow{2}{*}{Type} & \multirow{2}{*}{Method} & \multicolumn{5}{c}{ABIDE I $\rightarrow$ ABIDE II}  \\
    \cmidrule(lr){3-7} 
    & & Accuracy & Precision & Recall & F1-score & AUC \\
    \midrule
    
    \multirow{2}{*}{\parbox{3cm}{\centering Traditional FC}}
   & MLP
   & 55.7 & 53.4 & 89.9 & 67.0 & 64.5 \\
    & SVM      & 51.9 & 51.0 & 93.7 & 66.1 & 60.7 \\

   \midrule
 
    \multirow{5}{*}{\parbox{3cm}{\centering Graph Neural Networks}}
     & GCN
   & 52.7 & 51.6 & 87.0 & 64.8 & 60.2 \\
    & GAT
 & 56.7 & 53.9 & 93.3 & 68.3 & 66.0 \\
    & BrainGNN
  & 58.6 & 56.7 & 72.7 & 63.7 & 64.8 \\
     & FBNetGen
   & 49.8 & 49.4 & 16.0 & 24.1 & 47.1 \\
    & FROG
  & 51.7 & 51.2 & 73.1 & 60.2 & 53.7 \\
    \midrule

    \multirow{3}{*}{\parbox{3cm}{\centering  Graph Transformers}}
     & Graph Transformer
    & 53.8 & 52.1 & 95.8 & 67.5 & 56.1 \\
    & SAN&
  51.1 & 50.5 & \textbf{97.9} & 66.7 & 62.4 \\
    & Graphormer
   & 50.0 & 50.0 & 50.0 & 35.1 & 58.7 \\
    \midrule
     \multirow{3}{*}{\parbox{3cm}{\centering  Brain Transformers}}
       & BNT
 & 59.9 & 60.1 & 72.7 & 64.2 & 64.8 \\
    & ALTER
    & 58.8 & 56.9 & 59.9 & 63.8 & 64.8 \\
    & CAGT
    & 52.7 & 55.7 & 26.9 & 36.3 & 58.2 \\
    \midrule
    
    \multirow{1}{*}{Our Framework}
      
    & EBT &
    \textbf{61.8} & \textbf{64.7} & 73.9 & \textbf{69.0} &\textbf{ 68.0} \\
    \bottomrule
  \end{tabular}
  \label{tab:performance_comparison_abide2} 
  }
\end{table}

\begin{figure*}[htbp]
    \centering
    \includegraphics[width=0.98\textwidth]{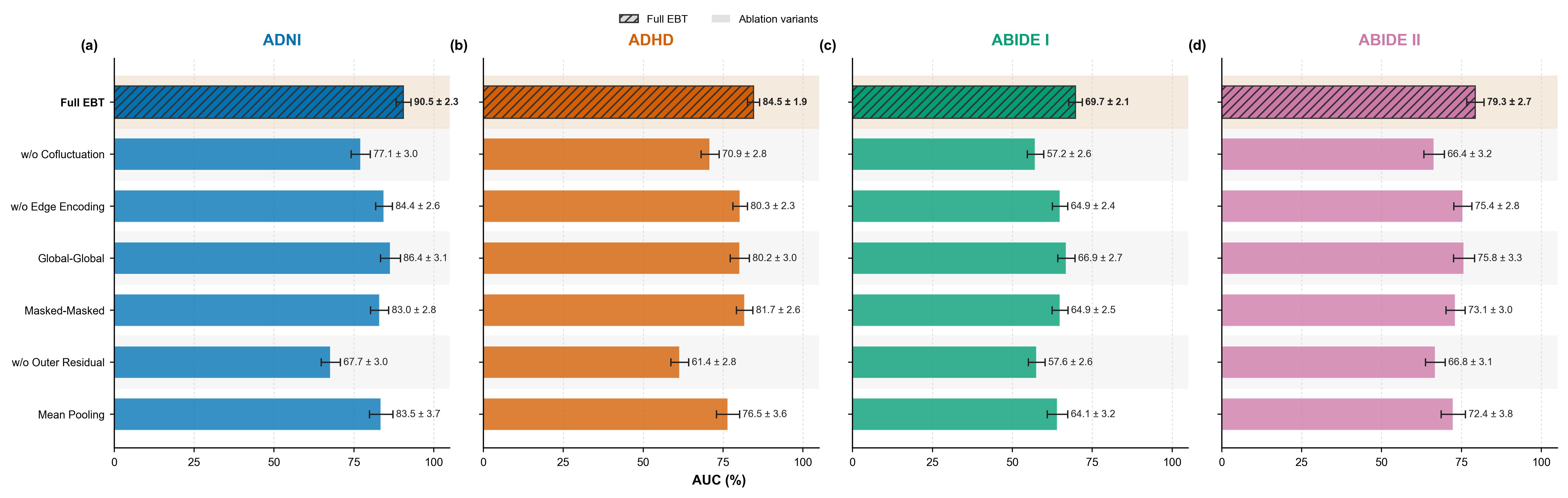}
    \caption{
    Ablation results of EBT on all datasets. Bars represent the mean AUC across five folds, and error bars indicate the standard deviations. Full EBT denotes the complete model, while other variants remove or replace individual components, including edge-wise temporal co-fluctuation representation, edge encoding, masked--global attention, outer residual connection, and Edge-OCRead. Global--Global and Masked--Masked replace the proposed attention scheme with purely global or masked attention, respectively. Mean Pooling replaces Edge-OCRead with simple mean aggregation.
    }
    \label{fig:ablation}
\end{figure*}
\subsection{External Validation}
To further evaluate cross-dataset generalizability, we conducted an external validation in which all models were trained and selected exclusively on ABIDE I and subsequently evaluated on the ABIDE II cohort without further model hyperparameter tuning. As shown in Table~\ref{tab:performance_comparison_abide2}, EBT achieved the highest accuracy, precision, F1-score, and AUC, reaching 61.8\%, 64.7\%, 69.0\% and 68.0\%, respectively. In terms of accuracy, EBT outperformed the strongest baseline, ALTER, by 3\%, while its AUC exceeded that of the best competing method, GAT, by 2\%. Although SAN obtained the highest recall of 97.9\% , its corresponding accuracy was only 51.1\% . The extremely high recall but near-chance accuracy observed for several baselines, including SVM, Graph Transformer, and SAN, indicates a strong tendency to predict the positive class, resulting in an imbalanced sensitivity--specificity trade-off. In contrast, EBT maintained a recall of 73.9\% and an F1-score of 69.0\% while achieving the best overall accuracy and discriminative ability. These results demonstrate that EBT provides a more balanced decision profile and retains stronger generalization performance under the distribution shift from ABIDE I to ABIDE II.

\subsection{Ablation Study}
To investigate the contribution of each component in EBT, we conducted ablation experiments on all four datasets. The variants include removing the edge-wise temporal co-fluctuation representation, edge encoding, and outer residual connection, replacing the sequential masked--global attention with Global--Global or Masked--Masked attention, and substituting Edge-OCRead with mean pooling. To provide a consistent and threshold-independent comparison across datasets, we report the AUC. As
shown in Fig.~\ref{fig:ablation}, removing any component consistently reduces the AUC across all four datasets, confirming the effectiveness of each design choice. Specifically, removing the edge-wise co-fluctuation representation causes substantial performance degradation, highlighting the importance of temporally resolved connectivity information beyond static FC. The performance decrease observed with Global--Global and Masked--Masked attention demonstrates that local topology-aware interactions and global edge dependencies provide complementary information. Furthermore, edge encoding, the outer residual connection,
and Edge-OCRead contribute to preserving edge-specific information and learning effective subject-level representations, with mean pooling showing inferior performance.
\subsection{Hyperparameter Analysis}
\begin{figure*}[t]
    \centering
    \includegraphics[width=0.98\textwidth]{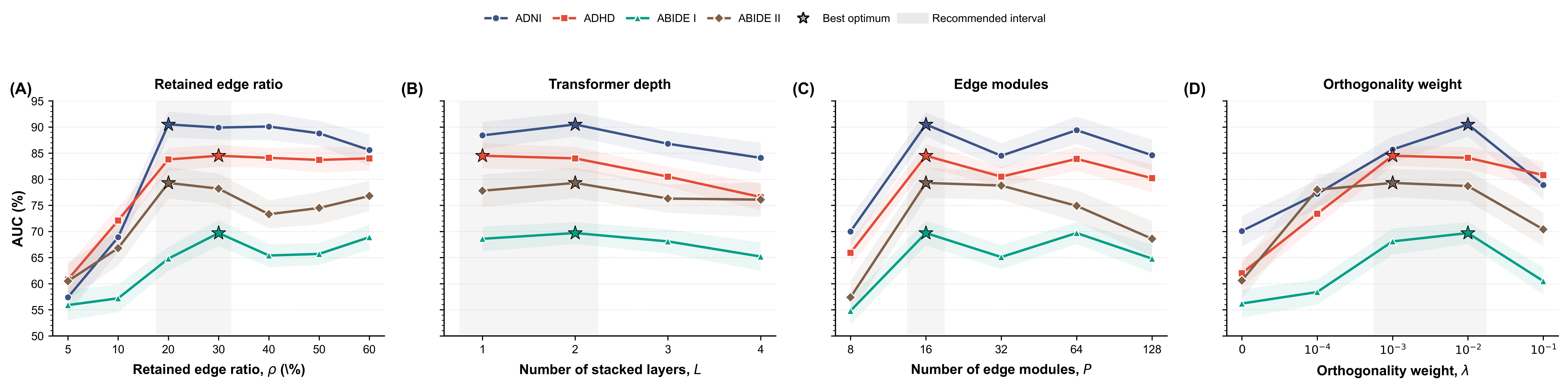}
    \caption{
    Hyperparameter sensitivity analysis of EBT on the ADNI, ADHD, ABIDE I,
    and ABIDE II datasets. 
    (A) Retained edge ratio \(\rho\);
    (B) number of stacked transformer layers \(L\);
    (C) number of latent edge modules \(P\); and
    (D) orthogonality regularization weight \(\lambda\).
    The curves represent the mean AUC across five folds, and the shaded
    regions indicate the corresponding standard deviations.
    Stars mark the dataset-specific optimal settings, while the gray regions
    indicate the recommended parameter intervals.
    }
    \label{fig:hyperparameter}
\end{figure*}
The sensitivity analysis of EBT with respect to the retained edge ratio
\(\rho\), the number of stacked transformer layers \(L\), the number of edge
modules \(P\), and the orthogonality weight \(\lambda\) is presented in
Fig.~\ref{fig:hyperparameter}. When \(\rho\) is too small, only a limited
number of functional connections are retained, resulting in insufficient
disease-related information; however, an excessively large \(\rho\)
introduces redundant and noisy connections. Therefore, EBT achieves more
stable performance when \(\rho\) is within \(20\%\)–\(30\%\). A shallow
transformer cannot adequately capture interactions among functional
connections, whereas excessive layers may lead to over-smoothing and
overfitting, with \(L=1\)–\(2\) generally providing the best performance.
Similarly, \(P\) should be sufficiently large to represent distinct
connectivity modules while avoiding redundant fragmentation, with the
optimal value observed at \(P=16\). For the orthogonality weight,
very small or large \(\lambda\) values respectively provide insufficient
module separation or impose excessive constraints, while stable performance
is achieved within \(10^{-3}\)–\(10^{-2}\). Overall, the optimal settings vary slightly across datasets but remain
within similar ranges, suggesting that EBT achieves relatively stable
performance under appropriate hyperparameter configurations across datasets.

\subsection{Model Interpretation}
\begin{figure}[h]
    \centering
    \includegraphics[scale=0.72]{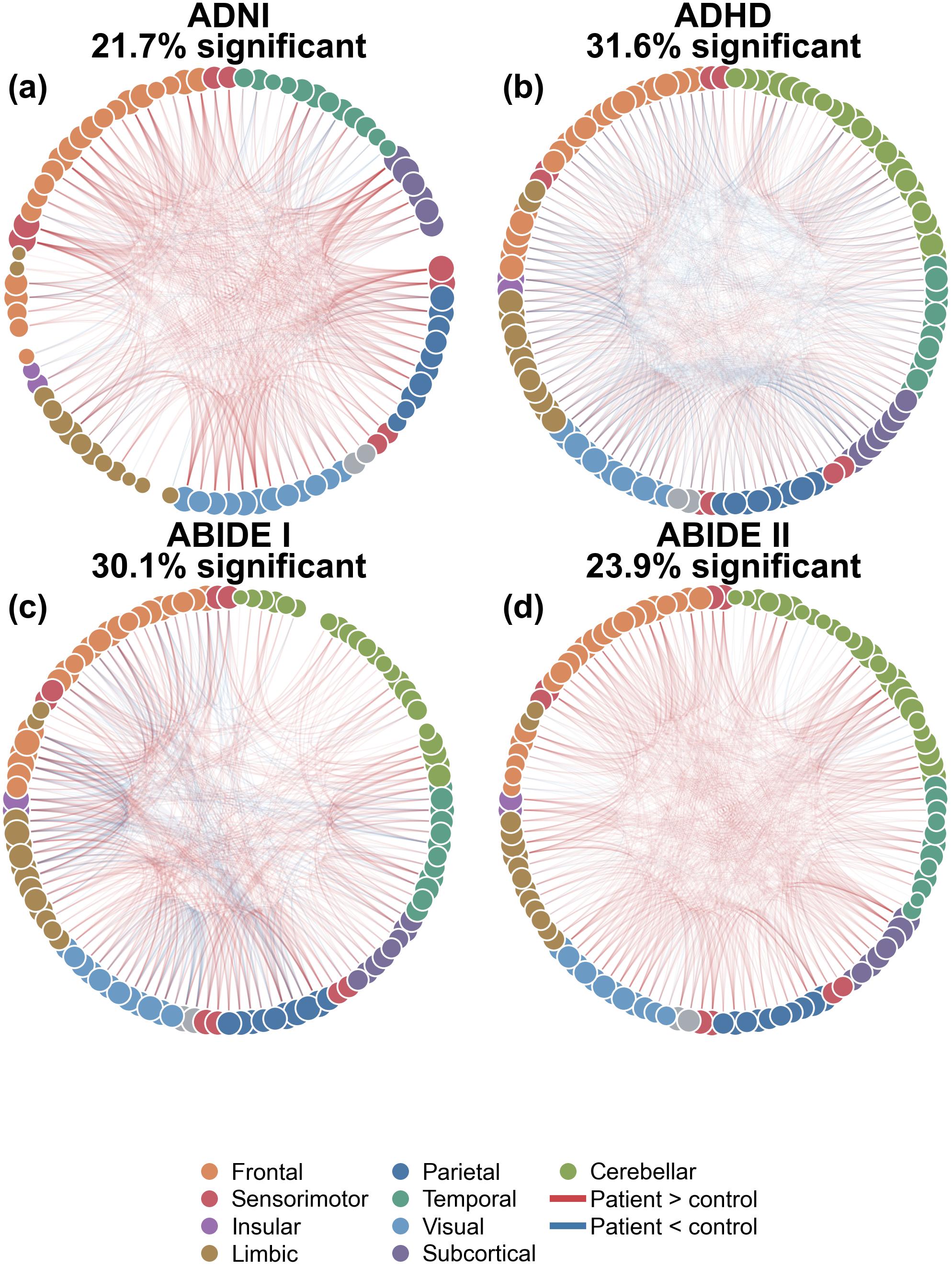}
    \caption{
    Statistical relevance and neuroanatomical distribution of the functional connections retained by EBT. Within each training dataset, the top \(30\%\) of connections were retained for ADHD and ABIDE I, while the top \(20\%\) were retained for ADNI and ABIDE II according to their group-difference statistics. The circular plots show the union of the selected connections that remained significant after BH-FDR correction in the corresponding training datasets.
    }
    \label{fig:significant_connectomes}
\end{figure}
\subsubsection{Interpretation of Selected Connections}
To examine whether the group-difference-based edge selection retained
statistically relevant functional connections, we calculated the proportion
of all connections that remained significant after BH-FDR correction
within each training dataset \cite{benjamini1995controlling}. As shown in
Fig.~\ref{fig:significant_connectomes}, the top \(20\%\) of connections
were selected for ADNI and ABIDE II, while the top \(30\%\) were selected
for ADHD and ABIDE I according to their group-difference statistics. Among
these connections, \(21.7\%\), \(31.6\%\), \(30.1\%\), and
\(23.9\%\) remained significant after multiple-comparison correction for
ADNI, ADHD, ABIDE I, and ABIDE II, respectively. These results indicate
that the statistical ranking strategy effectively enriches functional
connections with group-discriminative information, supporting its use for
constructing the edge-centric input graph.

The significant connections were distributed across brain regions
previously associated with the corresponding disorders. In ADNI, the
retained connections mainly involved the thalamus, supplementary motor
area, paracentral lobule, lingual gyrus, and precuneus, suggesting
alterations in thalamo-sensorimotor, default-mode, and visual-related
circuits \cite{greicius2004default}. In ADHD, the selected connections
involved the caudate nucleus, insula, cingulate cortex, calcarine cortex,
and cerebellum, which are consistent with previously reported
abnormalities in cortico-striatal, salience, visual, and cerebellar
networks \cite{bush2005functional}. For both ABIDE cohorts, significant
connections were frequently observed among the medial and orbitofrontal
cortices, insula, precuneus, supramarginal gyrus, thalamus, and cerebellum,
corresponding to alterations in default-mode, salience, frontoparietal,
and thalamocortical systems implicated in autism spectrum disorder
\cite{lau2019resting}. Overall, the statistical significance and neuroanatomical distribution of
the retained connections provide supporting evidence that the proposed
edge-selection strategy preserves biologically meaningful functional
connectivity patterns.

\subsubsection{Interpretation of Latent Connectivity Classes}

\begin{figure*}[t]
    \centering
    \includegraphics[width=0.97\textwidth]{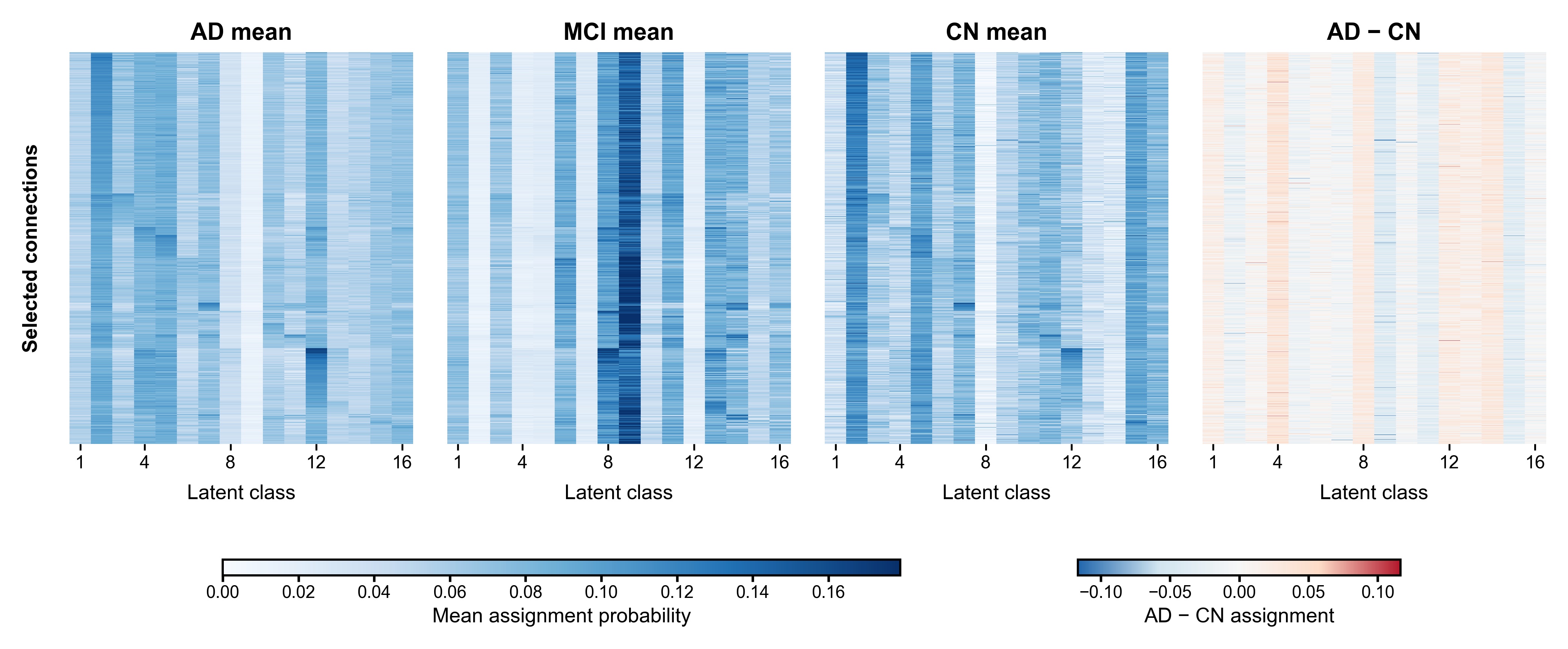}
    \caption{
Interpretation of latent connectivity classes learned by Edge-OCRead on the
ADNI dataset. The rows represent the selected functional connections and the
columns denote latent connectivity classes. The heatmaps show the averaged
edge-to-class assignment probabilities for AD, MCI, and CN groups, while the
right panel displays the assignment difference between AD and CN groups.
Different assignment patterns across groups indicate that Edge-OCRead learns
disease-associated higher-order organizations of functional connections.
}
    \label{fig:edge_ocread_assignment}
\end{figure*}
To further investigate whether the proposed Edge-OCRead module learns
meaningful higher-order organizations of functional connections, we analyzed
the learned edge-to-class assignment matrix
\(\mathbf{C}=\mathrm{softmax}(\mathbf{Z}\mathbf{W}_c)\).
Each element \(C_{kp}\) represents the probability that the \(k\)-th
functional connection is assigned to the \(p\)-th latent connectivity
class. Unlike the previous statistical analysis based on predefined
group-difference edges, this analysis directly characterizes the latent
connectivity structures learned by EBT.

For each subject, the assignment matrices were extracted from the
models and averaged within each diagnostic group. Taking the
ADNI as an example, Fig.~\ref{fig:edge_ocread_assignment} presents
the group-level edge-to-class assignment patterns for AD, MCI, and CN
subjects. Although the three groups share similar overall latent class
organizations, distinct assignment patterns can be observed across
diagnostic groups, indicating that Edge-OCRead captures group-dependent
variations in the organization of functional connections. The assignment
differences between AD and CN further suggest that disease-related changes
are reflected in the higher-order arrangement of connectivity patterns
rather than isolated individual connections.

Furthermore, within each latent connectivity class, multiple functional
connections exhibit consistently high assignment probabilities, suggesting
that the learned classes represent coherent connectivity organizations
instead of arbitrary partitions of edges. These results demonstrate that
Edge-OCRead can automatically discover biologically meaningful latent
connectivity structures and provide an interpretable edge-centric
representation for characterizing functional network alterations.

\subsubsection{Biomarker Interpretation}
\begin{figure*}[t]
    \centering
    \includegraphics[width=\textwidth]{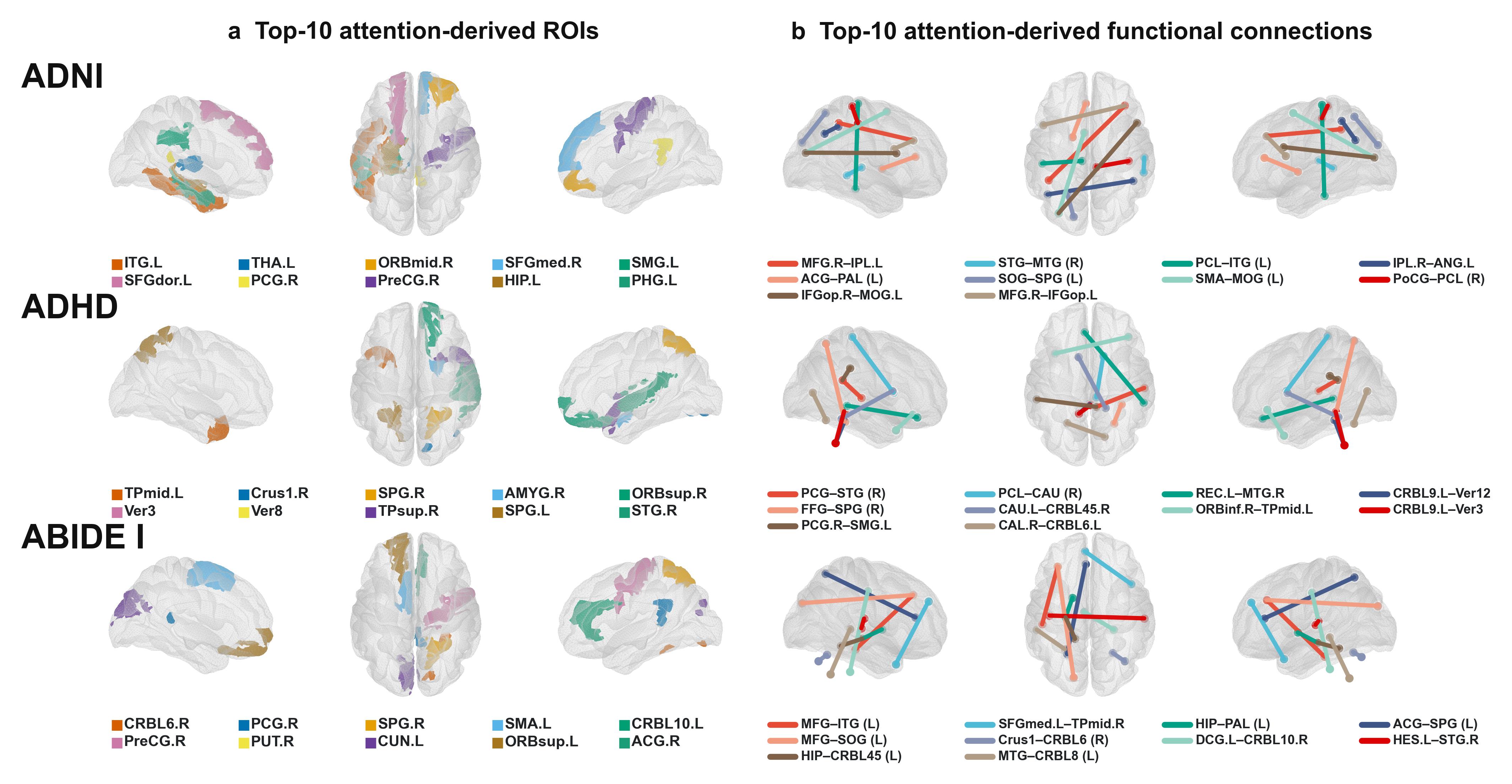}
    \caption{
Attention-based interpretation of EBT on representative neuroimaging
datasets. The top-10 attention-derived brain regions (left) and functional
connections (right) are shown for ADNI, ADHD, and ABIDE I. Node colors denote
anatomical regions, and edge colors represent different functional
connections. The results demonstrate that EBT focuses on disorder-relevant
brain regions and connectivity patterns rather than uniformly distributed
connections across the whole brain.
}
    \label{fig:attention_biomarker}
\end{figure*}
To interpret the discriminative patterns learned by EBT, we analyzed
the attention weights from the EBT transformer layers to identify
the most influential brain regions and functional connections. Since each
node in the line graph represents a functional connection, the attention
weights quantify the contribution of individual connections during
information aggregation. For each subject, the attention scores were
aggregated to obtain the importance of functional connections and their
associated brain regions. The subject-level importance scores were then
averaged across individuals to identify group-level discriminative patterns.

Figure~\ref{fig:attention_biomarker} presents the top-10 attention-derived
regions and functional connections for ADNI, ADHD, and ABIDE I datasets.
The identified regions and connections showed disorder-specific
distributions rather than random localization. In ADNI, the important
regions mainly involved the inferior temporal gyrus, thalamus, precuneus,
hippocampus, parahippocampal gyrus, and supplementary motor regions. These
regions are closely related to memory processing, default-mode network
organization, and thalamocortical communication, which have been repeatedly
implicated in Alzheimer's disease-related functional alterations
\cite{greicius2004default,wu2022abnormal}. The highly weighted connections
also involved interactions among temporal, parietal, and subcortical
regions, suggesting that EBT captured disrupted large-scale connectivity
patterns associated with cognitive decline. For ADHD, the attention-derived regions were mainly distributed across the
temporal, frontal, parietal, and cerebellar areas, including the superior
temporal gyrus, orbitofrontal regions, and cerebellar structures. The
identified functional connections primarily involved cortico-striatal,
frontoparietal, and cerebellar circuits, consistent with previous findings
that ADHD is associated with altered executive-control, salience, and
cortico-striatal connectivity patterns
\cite{cao2006abnormal,mostert2018similar}. For ABIDE I, the important regions and connections were mainly located in
the default-mode, frontoparietal, and cerebellar systems, including the
precuneus, posterior cingulate-related regions, frontal areas, and
cerebellar subregions. These findings agree with previous neuroimaging
studies reporting atypical functional connectivity involving
default-mode, social cognition, and thalamocortical networks in autism
spectrum disorder
\cite{yerys2015default,woodward2017thalamocortical}.

Overall, the attention-derived regions and connections provide an
interpretable view of how EBT makes predictions from edge-centric functional
representations. Importantly, these findings should be regarded as
model-derived discriminative patterns rather than definitive clinical
biomarkers. The consistency between the identified patterns and previously
reported disease-related networks suggests that the proposed model captures
meaningful neurobiological information from functional connectivity
organization.

\section{Discussion}
In this study, we proposed an Edge-centric  Brain Transformer (EBT) for
rs-fMRI-based brain disorder diagnosis by redefining the fundamental
learning unit from brain regions to functional connections. Unlike
conventional node-centric graph learning approaches, EBT explicitly models
functional connections as edge tokens and captures their interactions
through a line graph-based transformer architecture. This design provides
a different perspective for functional connectome analysis by focusing on
the organization of connectivity patterns rather than isolated regional
features.

The effectiveness of EBT can be attributed to several aspects. First, the
edge-wise temporal co-fluctuation representation enables the model to
capture dynamic information embedded in functional interactions. Static
functional connectivity summarizes temporal correlations into a single
value, which may overlook transient coordination patterns between brain
regions. By preserving temporal co-fluctuation characteristics, EBT retains
additional discriminative information that may reflect subtle alterations
in functional communication associated with brain disorders. Second, the line graph formulation allows direct modeling of
connection-to-connection relationships. Existing graph-based brain network
methods mainly treat regions as nodes and use functional connections as
edges for information propagation. However, disease-related abnormalities
often emerge from distributed changes involving multiple interacting
connections. By representing functional connections as nodes in a line
graph, EBT can explicitly learn higher-order connectivity organizations.
The sequential masked and global attention mechanism further integrates
local topology constraints with long-range functional dependencies. Third, Edge-OCRead provides an interpretable representation learning
strategy by discovering latent connectivity modules instead of relying on
simple global pooling. The learned edge assignment patterns revealed
group-dependent organization of functional connections, suggesting that
the model captures structured variations in higher-order connectivity
patterns across diagnostic groups.

Beyond predictive performance, the interpretability analyses provided
additional evidence for the biological relevance of the learned representations. The selected connections showed significant group
differences after multiple-comparison correction, while attention-derived
regions and connections were consistent with previously reported disorder-related networks. These findings indicate that EBT learns
meaningful connectivity representations rather than relying on arbitrary statistical patterns. Despite these advantages, several limitations remain. First, although EBT was evaluated on four independent cohorts, larger multi-site datasets are needed to further validate its scalability and generalization ability. Second, the current framework performs statistical edge selection before transformer encoding to reduce computational complexity. Future work could
explore adaptive end-to-end edge discovery strategies that jointly optimize
connection selection and representation learning. Third, the identified attention patterns and latent connectivity modules should be interpreted as model-derived discriminative features rather than direct clinical biomarkers. Additional longitudinal and multimodal studies are required to evaluate their potential clinical utility.

\section{Conclusion}

This study introduced EBT, an edge-centric brain transformer framework for
learning functional connectivity representations from rs-fMRI data. By
combining temporal co-fluctuation modeling, line graph-based connection
interaction learning, and interpretable connectivity module discovery, EBT
provides a new approach for characterizing higher-order functional network
organization.

Experiments across four independent neuroimaging cohorts demonstrated the
effectiveness of the proposed framework, while interpretability analyses
showed that the learned representations captured structured and
biologically relevant connectivity patterns. These results highlight the
potential of edge-centric learning as a complementary paradigm to
traditional region-based brain network analysis.

Future studies will focus on scaling EBT to larger multi-site datasets,
developing adaptive connectivity discovery strategies, and integrating
multimodal information to further improve robustness and clinical
translation.

\bibliographystyle{IEEEtran}
\bibliography{reference}

\end{document}